\documentclass[%
 reprint,
 amsmath,amssymb,
 aps,
]{revtex4-1}

\usepackage{graphicx}
\usepackage{dcolumn}
\usepackage{multirow} 
\usepackage{colortbl} 
\definecolor{kugray5}{RGB}{224,224,224}
\usepackage{bm}

\begin{document}

\preprint{APS/123-QED}

\title{CEI: a new indicator measuring City Commercial Credit Risk initiated in China}

\author{Ruonan Lin$^{1}$}
 \email{Correspondence to: linruonan@gmail.com}
 \author{Yi Gu$^{1,2}$}%
\affiliation{%
 $^{1}$Department of Applied Mathematics and Statistics, Stony Brook University, Stony Brook, NY 11794, USA\\
 $^{2}$Department of Physics and Astronomy, Stony Brook University, Stony Brook, NY 11794, USA
}%

\date{\today}

\begin{abstract}
Aiming at quantifying and evaluating the regional commercial environment along with the level of economic development among cities in mainland China, the concept of China City Commercial Environment Credit Index(CEI) was first introduced and established in 2010. In this manuscript, a historical review and detailed introduction of CEI is included, followed by statistical studies. In particular, an independent statiscial cross-check for the existing CEI-2012 is performed and significant factors that play the most influential roles are discussed.

\end{abstract}

\pacs{Valid PACS appear here}
\maketitle


\section{\label{sec:level1} History Review and Paper Structure}

China, one of the fastest growing economies in the world, has always been attracting worldwide attention in the past two decades.  After a period of stable and rapid development, China has entered a new era of credit economy. By the end of last century, China initiated to build its own social credit system, of which the scale and importance are almost comparable to the social security system in the United States. In just over one decade of time, this multi-facet system, from its coming into existence, has already exhibited increasing power of steering China's economic and social development.  

Social credit system functions to increase credit launch, control credit risk and enhance credit awareness\cite{Lin2003SocialCreditSys}. Chinese government expects the credit system to act in avoiding, controling and transferring credit risk as well as to establish a ``credit punishment mechanism", based upon which it is also anticipated to help regulate market behaviors and nurture new business philosophy. Because of the ``punishment mechanism", social credit system becomes the only feasible solution to dishonest deed and discreditable behaviors, and thus it is cruicial to the construction and regulation of market within framework of credit economy. 

Not until recently did China realize that its government is in urgent need of creating a series of economic indices to help evaluate the development level of urban social credit system and to quantify the commercial credit risk. To this end, China City Commercial Environment Credit Index(China City CEI, or CEI for short) was introduced by Chinese Academy of Management Science in 2010. Each participating city is assigned a numerical value(within 30-100), which is carefully calculated via a specific statistical model. Taking into account multiple commercial factors, the CEI is expected to reflect the status quo of local credit system, and it serves as a tool to evaluate commercial environment. Since ``city" is the unit entity in the system, CEI is calculated on a city-wide basis. As an index computed and tested annually, CEI helps to build the whole social credit system in that it quantifies all the relavant commercial factors in the realm of economics and sociology, and it makes the credit system more concrete, measureable and comparable. 

A group of CEI initiators has tested this economic index among 248 cities in China, and researchers have been working diligently on its improvement. After the first two release periods(CEI for year 2011, 2012), both the central and local government in China have paid great attention to it. The CEI will surely influence the government policy-making as one of the most important economic indices to refer to in future.

In this paper we include a relatively detailed introduction of CEI in the second part, including factors involved that are important to construct CEI. In the third part, on top of the existing categorical CEI-2012(i.e., the CEI's based on each index from the first index class, see later section), we present an independent statistical cross-check using Principal Component Analysis(PCA) techniques, from which the most influntial factors in CEI are extracted. A paired test is also performed, and the result from the test indicates that CEI is indeed a convincing tool to evaluate the status quo of commercial environment among cities in mainland China. 
In the last part of the paper, the importance and future influence of CEI will be discussed. 
 
\section{\label{sec:level1} Introduction}

CEI, an unprecedented example of applying theories of social credit system into practice, has been regarded as a remarkable milestone over the course of city development history in China.  As a comprehensive economic index, CEI is being used to measure and evaluate the level of credit growth, the trading levels in credit market, the goodness of a market, and the overall level of social credit awareness. It enables us to analyze the complicated social economics phenomena. To be more specific, CEI reflects, but not limited to, the following regional feature of commercial environment~\cite{CEIreport}:
\begin{itemize}
\item The activities of urban credit economics, for example, the variation of credit instrument launch and enterprise spot sale on credit
\item The fluctuation and operational level of urban credit economics
\item The scale of urban credit transactions
\item The risk level of credit transactions in local market
\item The establishement of risk by-law among local enterprises
\item The precautious steps that local enterprises take to avoid, control and transfer credit risk
\item To what extent the local market environment influences local enterprise's good standing and development
\end{itemize}

\subsection{\label{sec:level1} Evaluation Index System}

Given the resources we have for collecting reliable and utilizable credit information(data) in China, CEI is established upon the following multiple commercial factors, namely credit launch, enterprise risk managment, construction of credit reporting system, credit monitoring services from local government, the odds of dishonesty deed and discreditable behaviors, the build-up of credit reward and punishment system. These factors comprise the \textit{first-class index} in the China city commercial environment evaluation index system. 

Most of the indices in first-class reflect sociological aspects of group behavior, and therefore they are quite abstract and hard to directly quantify. For this reason we introduce \textit{second-class index}, which are sligtly more concrete. The \textit{thrid-class index} is more or less the embodiment or concretization of the first two class indices. All the three class indices comprise the entire China city commercial environment evaluation index system. The entire index system is included in Table.\ref{tab:table_indexSystem}\cite{Bluebook2012CEI}.

\begin{table*}
\def\arraystretch{1.3}
\begin{tabular}{ |l|l|l| }
\hline
{\mbox{ } \textbf{First-class Index}}  & { \mbox{ } \mbox{ } \mbox{ } \mbox{  } \mbox{ } \textbf{Second-class Index }}  & { \mbox{ } \mbox{ } \mbox{ }  \mbox{ } \mbox{ } \mbox{ } \mbox{ } \mbox{ } \mbox{ } \mbox{ } \mbox{ } \mbox{ } \mbox{     }\textbf{Third-class Index }} \\
\hline\hline
\multirow{7}{*}{\mbox{ } \mbox{ } \mbox{ } credit launch} & & \mbox{     } the proportion of consumer credit balances in GDP \\
\cline{3-3}
 &  \mbox{  } \mbox{ }\mbox{ }\mbox{ } credit instrument launch &\mbox{     } the proportion of credit loan from small and medium-sized \\
  &  & enterprises in overall credit loan  \\
 \cline{2-3}
 &   & \mbox{     }the proportion of credit sales contracts in overall sales contracts\\
 \cline{3-3}
 &  \mbox{  }\mbox{ }\mbox{ }\mbox{ }\mbox{ } commercial credit sales& \mbox{     } the proportion of gross credit sales in GDP\\ 
 \cline{3-3}
 &  & \mbox{     }the proportion of agreement-honored credit sales in gross  \\
 & & credit sales  \\ \hline

\multirow{7}{*} & \mbox{   } establishment of credit manage-  & \mbox{     } the proportion of enterprise with credit management dept. \\
 \cline{3-3}
 &\mbox{ }  ment department in enterprise &\mbox{     } the client enterprises of large credit reporting agencies  \\
 \cline{2-3}
 \mbox{ }  \mbox{ } \mbox{ } enterprise risk&  & \mbox{ }  the proportion of credit management licensee in all employees  \\
 \cline{3-3}
 \mbox{  } \mbox{ } \mbox{ } management & \mbox{     } \mbox{ }\mbox{ } risk management talents  & \mbox{ } the employees that attended risk management training\\ 
 \cline{3-3}
  &  & \mbox{ } risk management major graduates from local institutes \\ 
  \cline{2-3}
 & \mbox{     }\mbox{ } \mbox{ }\mbox{ }\mbox{ } enterprise financial risk & Themis$^{\copyright}$financial risk early-warning index of  the listed companies \\ \hline
 \multirow{6}{*}&\mbox{ } development of credit-reporting& \mbox{ } the variety/completeness of types of credit reporting agencies  \\
\cline{3-3}
 \mbox{ }  \mbox{ } construction of  &\mbox{  } \mbox{ }  \mbox{ }\mbox{ }\mbox{ }\mbox{ }\mbox{ } \mbox{ }\mbox{ }\mbox{ }\mbox{ }\mbox{ } industry & \mbox{ } the number of credit reporting agencies \\
 \cline{2-3}
\mbox{ }  \mbox{ } credit reporting & credit info reports, related products   & \mbox{ } the variety of credit products and services provided  \\
 \cline{2-3}
\mbox{ } \mbox{ } \mbox{ }  \mbox{ }  \mbox{ } system  & \mbox{  } \mbox{ }\mbox{ } utilization of public credit  & \mbox{ } the proportion of enterprises whose credit info are consulted \\ 
 \cline{3-3}
 & \mbox{ } reporting system (on a yearly basis)  & \mbox{ } the proportion of individuals whose credit info are consulted   \\ 
  \cline{3-3} \hline 
 
 \multirow{6}{*} &\mbox{ } establishment of monitoring dept. & \mbox{ } local governments with credit monitoring department \\
\cline{2-3}
 &   & \mbox{ } the well-being and upgrade status of credit infrastructure \\
 \cline{3-3}
 \mbox{ } \mbox{ } credit monitoring & \mbox{ } \mbox{ } \mbox{ } \mbox{ } credit infrastructure & \mbox{ } whether ``blacklist system" is set up in monitoring dept.\\
 \cline{2-3}
\mbox{ } \mbox{ } from government &  & \mbox{ } the amount of by-law or regulations established\\ 
 \cline{3-3}
 & establishment and implementation of  & \mbox{ } the amount of credit standards established \\ 
  \cline{3-3}
 & \mbox{ } \mbox{ }the by-law of credit transaction  & \mbox{ } the implementation of national credit standards\\ \hline 
 
\multirow{6}{*}  & & \mbox{ } the number of economic crimes of swindling \\
\cline{3-3}
 &  \mbox{   } \mbox{   }\mbox{   }\mbox{   } \mbox{   } \mbox{   } economic law cases & \mbox{ } the amount of money involved in swindling crimes \\ \cline{3-3}
 \mbox{ } \mbox{ }\mbox{ } \mbox{ } \mbox{ } discredit &  & \mbox{ } crimes of producing and selling fake and inferior goods \\
 \cline{2-3}
  \mbox{ }\mbox{ }\mbox{ }\mbox{ }dishonesty deed &  & \mbox{ } major discredit criminal cases \\
 \cline{3-3}
  &  \mbox{ } \mbox{ }  major discredit criminal cases  & \mbox{ } bribe and corruption crimes  \\ 
 \cline{3-3}  \hline
 
 \multirow{3}{*}  &\mbox{ } \mbox{ }  the build-up of credit system & \mbox{ } whether personnel identified to take over the system build-up  \\
 \cline{2-3}
 \mbox{ } \mbox{ } credit education  &  & \mbox{ } number of events on credit education \\
 \cline{3-3}
  &  \mbox{ } the status quo of credit education   & \mbox{ } whether credit education info availabe online(via web,vedio...) \\ 
 \cline{3-3}  \hline
 
\mbox{ } corporate experience & \mbox{ } satisfaction with credit environment & \mbox{ } questionnaire that covers over 30 categories  \\
 \hline
\end{tabular}
\caption{\label{tab:table_indexSystem} China City Commercial Environment Evaluation Index System}
\end{table*}

\subsection{\label{sec:level1} Data Sources and Data Processing}

CEI is a retrospective study of city credit environment, that is, it is calculated based upon data collected in the prior fiscal year. For example, the CEI released in July of 2012 utilizes the data collected between Apr.1$^{st}$, 2011 and Mar.31$^{st}$, 2012. By convention of economic index, CEI reflects the feature(commercial environment of a city) of last year. As information is collected every year, the cumulative data over a long period of time allow us to use regression analysis technique to predict future behaviors(i.e., city credit environment).

To ensure reliability of data and autority of CEI, credit information is collected simultanously in all participating China cities according to stringent statistical model requirements. The CEI research group clearly defines each index and its corresponding way of collecting and processing data. The standardized procedure of data-processing is universal among all participating cities so as to guarantee cross-city consistency and comparability.

The data source selection is also very cautious and demanding. For CEI-2012, the collected credit information only comes from the following sources: (1) major credit reporting agencies; (2) government statistics; (3) credit monitoring authorities; (4) indirect but authoritative data; for example, CEI research group utilizes Themis$^\copyright$ financial risk early-warning index and its evaluation results to assess enterprise financial risk measures\cite{Pu2011Themis}; (5) questionnaires conducted by authorized research and development unit. 

There are dozens of factors involved in modeling CEI(listed in Table.\ref{tab:table_indexSystem}), therefore CEI calculation requires a \textit{comprehensive evaluation method}. This method uses multiple indices to evaluate multiple participating cities, and it is also named \textit{multi-index comprehensive evaluation method}. The underlying idea is to transform multiple indices into one index that could reflect the comprehensive situation, and we use that index to evaluate each participating city.

Generally speaking, there are two major steps that are indispensable in the evaluation process\cite{Xu2004StatIndex}: \\

(a) quantify credit information and pretreat indices of different units so as to remove the influence of non-uniform dimension\\

(b) calculate comprehensive evaluation index(or comprehensive evaluation score) based on the post-treatment index values. If linear weighting method is applied to pick up contributions from all indices, a weighting factor should be assigned for each index in certain index class.\\

Using multi-objective programming principles and the efficacy coefficient method, one is able to find upper and lower limit of each single index, and we further transform each single index into comparable evaluation score, i.e., single index evaluation value. Let $X_i$ denotes the raw value of index $i$, and $X^{i}_{min}$($X^{i}_{max}$) be the minimum(maximum) raw value of index $i$. $Z_i$ is the dimensionless evaluation score. For any ``positive index"(the index whose higher value is favored), the usual efficacy coefficient can be calculated:
\begin{eqnarray}
Z_i=\frac{x_i-x_{min}^{i}}{x_{max}^{i}-x_{min}^{i}} \times 100
\label{eq:one}.
\end{eqnarray} 
The modified efficacy coefficient can be found in Eq.(\ref{eq:two})
\begin{eqnarray}
Z_i=\frac{x_i-x_{min}^{i}}{x_{max}^{i}-x_{min}^{i}} \times 40+60
\label{eq:two}.
\end{eqnarray} 

After obtaining all dimensionless evaluation values(each corresponds to one index), the CEI research group uses linear weighted method to calculate overall CEI,
\begin{eqnarray}
CEI=\frac{\sum_{i=1} Z_i W_i}{\sum_{i=1} W_i}
\label{eq:three}.
\end{eqnarray} 
where in Eq.(\ref{eq:three}), $W_i$ is the weight for index $i$. Similar to the group decision making process, these weights are determined via Analytic Hierarchy Process(AHP) in subjective weighting method to minimize the influence of subjectivity.

\section{\label{sec:level1} Statistical Studies}

We perform a set of completely independent statistical studies as a cross-check to the CEI released in year 2012 using Principal Component Analysis(PCA) technique.

PCA is a statistical method that uses orthogonal transformations to convert a set of possibly correlated variables into a set of linearly uncorrelated variables. These uncorrelated variables are called \textit{principal components}. Such transformation is defined in a way that the first principal component accounts for as much of the variability in data as possible and could be used as a comprehensive strength indicator. The dependent variable in PCA is not an actual existing variable, but a variable defined by the analyst. In our studies, the original independent variables are the official categorical CEI's(or city rankings by categorical CEI) based on six of the indices from first class: credit launch, enterprise risk management, construction of credit reporting system, credit monitoring at government level, discredit/dishonesty, credit education. The dependent variable is the overall City Commercial Environment Credit Index (CEI).

\begin{table}
\begin{ruledtabular}
\begin{tabular}{lcccccc}
\textrm{ }&
\textrm{LCH}&
\textrm{ETP}&
\textrm{SYS}&
\textrm{GOV}&
\textrm{DIS}&
\textrm{EDU}\\
\colrule
LCH & 1.0000 & 0.1576 & 0.1933 & 0.1539 & 0.0849 &0.1654\\
ETP & 0.1576 & 1.0000 & -0.0149 & -0.1805 & -0.1170 & -0.2306 \\
SYS & 0.1933 & -0.0149 & 1.0000 & 0.3502 & 0.2343 & 0.4114 \\
GOV& 0.1539 & -0.1805 & 0.3502 & 1.0000& 0.2306 & 0.4438 \\
DIS & 0.0849 & -0.1170 & 0.2343 & 0.2306 & 1.0000 & 0.0847 \\
EDU& 0.1654 & -0.2306 & 0.4114 & 0.4438 & 0.0847 & 1.0000\\ 
\end{tabular}
\end{ruledtabular}
\caption{\label{tab:table2} covariance matrix for the six observables(indices) in first class. LCH: credit launch, ETP: enterprise risk management, SYS: construction of credit reporting system, GOV: credit monitoring at government level, DIS: discredit/dishonesty and EDU: credit education status}
\end{table}

\begin{table}[h!]
\begin{ruledtabular}
\begin{tabular}{lccccc}
\textrm{ }&
\textrm{Eigenvalue}&
\textrm{Difference}&
\textrm{Proportion}&
\textrm{Cumulative}\\
\colrule
1  &  2.05339008   & 0.86295359    &    0.3422    &    0.3422 \\
2  & 1.19043649   &  0.25762628    &   0.1984     &   0.5406 \\
3  &  0.93281021   & 0.19024314    &  0.1555      &  0.6961 \\
4  &  0.74256708   & 0.13428583    &  0.1238      &  0.8199 \\
5  &  0.60828125   &  0.13576637   &  0.1014      &  0.9212 \\
6  &  0.47251488   &                        & 0.0788       &  1.0000
\end{tabular}
\end{ruledtabular}
\caption{\label{tab:table3} eigenvalues of covariance matrix in Table.\ref{tab:table2}. The $i^{th}$ row corresponds to the $i^{th}$ eigenvector(pincipal component). ``Proportion" denotes the proportion(or percentage) of variance in data specific eigenvector could account for, and ``Cumulative" denotes cumulative proportion of variance in data if all eigenvectors up to current one(including current one) are included.}
\end{table}

\begin{table}[h!]
\begin{ruledtabular}
\begin{tabular}{lcccccc}
\textrm{ }&
\textrm{Y1}&
\textrm{Y2}&
\textrm{Y3}&
\textrm{Y4}&
\textrm{Y5}&
\textrm{Y6}\\
\colrule
LCH   &   0.244189  &    0.642825  &    -.054923  &    -.699277  &    -.151514  &    -.110388 \\
ETP    &  -.210139   &   0.728475   &   0.044816   &   0.473807  &    0.333973   &   0.295171 \\
SYS    &  0.496945   &   0.199181   &   0.008052   &   0.531121  &    -.488440   &   -.438913 \\
GOV   &   0.526149  &    -.075805  &    -.070938  &    -.000772 &     0.783788  &    -.313151 \\
DIS    &  0.310926    &  -.048091   &   0.899320    &  -.050786   &   -.012501    &  0.299165 \\
EDU   &   0.525120   &   -.091473  &    -.425559   &   0.042999 &     -.111586   &   0.721443 \\
\end{tabular}
\end{ruledtabular}
\caption{\label{tab:table4} eigenvectors of covariance matrix in Table.\ref{tab:table2}. Variable abbr.'s follow the same rule as in Table.\ref{tab:table2}}
\end{table}

The maximum number of principal components can be same as number of independent variables. The first principal component reflects the ``overall strength of response". The remaining principal components could be automatically calculated, and they may not have practical meaning. Each succedent component is required to have the highest possible variance under given constraint, which is orthogonal to the preceding components.

The structural formula for the principal component is expressed as:
\begin{eqnarray}
Z=\sum_i^{p} a_i u_i
\label{eq:four}.
\end{eqnarray} 
where $p$ is the number of original independent variables. $a_i$ is the coefficient reflecting the degree of influence that $i^{th}$ variable casts on the principal component. $u_i$ is the standardized value of variable $i$. 

The covariance matrix is calculated after each evaluation score is normalized because PCA is sensitive to the relative scale of the variables. The results can be found in Table.\ref{tab:table2}. The eigenvalues and eigenvectors of the covariance matrix can be found in Table.\ref{tab:table3} and \ref{tab:table4} respectively.

From Table.\ref{tab:table3}, we notice the largest eigenvalue $2.0534$ corresponds to the first principal component(Y1). The eigenvector Y1 returns coefficients for the involvded terms in the first principal component. Therefore the evaluation score based on the first principal component(denoted as Z1) can be expressed in Eq.(\ref{eq:five}):
\begin{eqnarray}
Z1=&0.244189*LCH-0.210139*ETP \nonumber \\
      &+0.496945*SYS +0.526149*GOV \nonumber \\
      &+0.310926*DIS
+0.525120*EDU
\label{eq:five}.
\end{eqnarray}
where LCH(our original variable) is the official city rankings by categorical CEI based on the factor of credit launch. Similarly for others.

The expressions for other principal components in terms of six index values can be easily achieved in the same manner. The principal component score($Z_i$) can be calculated accordingly. 

One possible criterion for determining whether to keep or discard a component is the proportion(or percentage) of variance in data such component accounts for. For example, you may decide to keep any component that accounts for at least 15\% or 20\% of the total variance in your model and drop the others. The proportion is defined and calculated using the following simple formula:
\begin{eqnarray}
\mbox{proportion} = \frac{\mbox{eigenvalue for the component of interest}}{\mbox{total eigenvalues of the correlation matrix}}\, \nonumber \\
\label{eq:six}.
\end{eqnarray}

In principal component analysis, the ``total eigenvalues of the correlation matrix" is qual to the total number of variables being analyzed, because each variable contributes one unit of variance to the analysis. Therefore, the larger the eigenvalue of the component is, the greater the proportion of variance in data that component accounts for. For example, the first pincipal component, with eigenvalue of 2.0534, accounts for approximately 2.0534/6 $\sim$ 34\% of the information in data. The second pincipal component comes next, which accounts for $\sim$19\% of variance in data.  

\begin{figure}[htbp]
\includegraphics[scale=0.22]{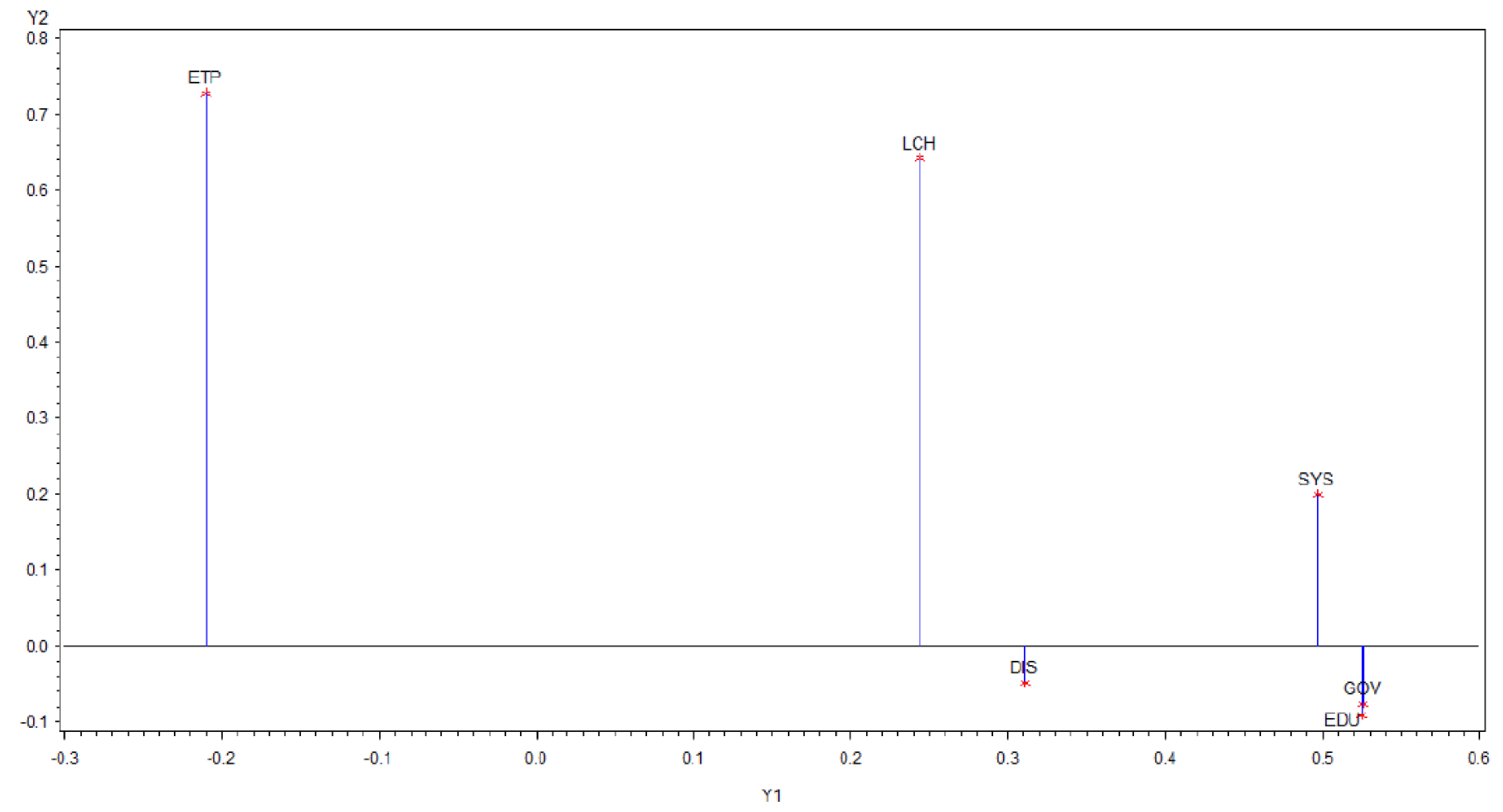}
\caption{\label{fig:index_rank_b} Coefficients of the six indices involved in this study for the first two pincipal components Y1, Y2. The horizontal axis is for Y1, and vertical axis is for Y2.}
\end{figure}

We also plot the eigenvectors of all six indices for the first two principal components Y1 and Y2 in Fig.\ref{fig:index_rank_b}. Since the first pincipal component weights almost twice as much as that of the second pincipal component, the most influential indices are determined by the coefficient of each piece in Y1. Within Y1, GOV(credit monitoring at government level) and EDU(credit education status) contribute roughly equal amount, hence they are the most influential indices in modeling overall CEI.

\begin{figure*}[htbp]
\includegraphics[width=170mm]{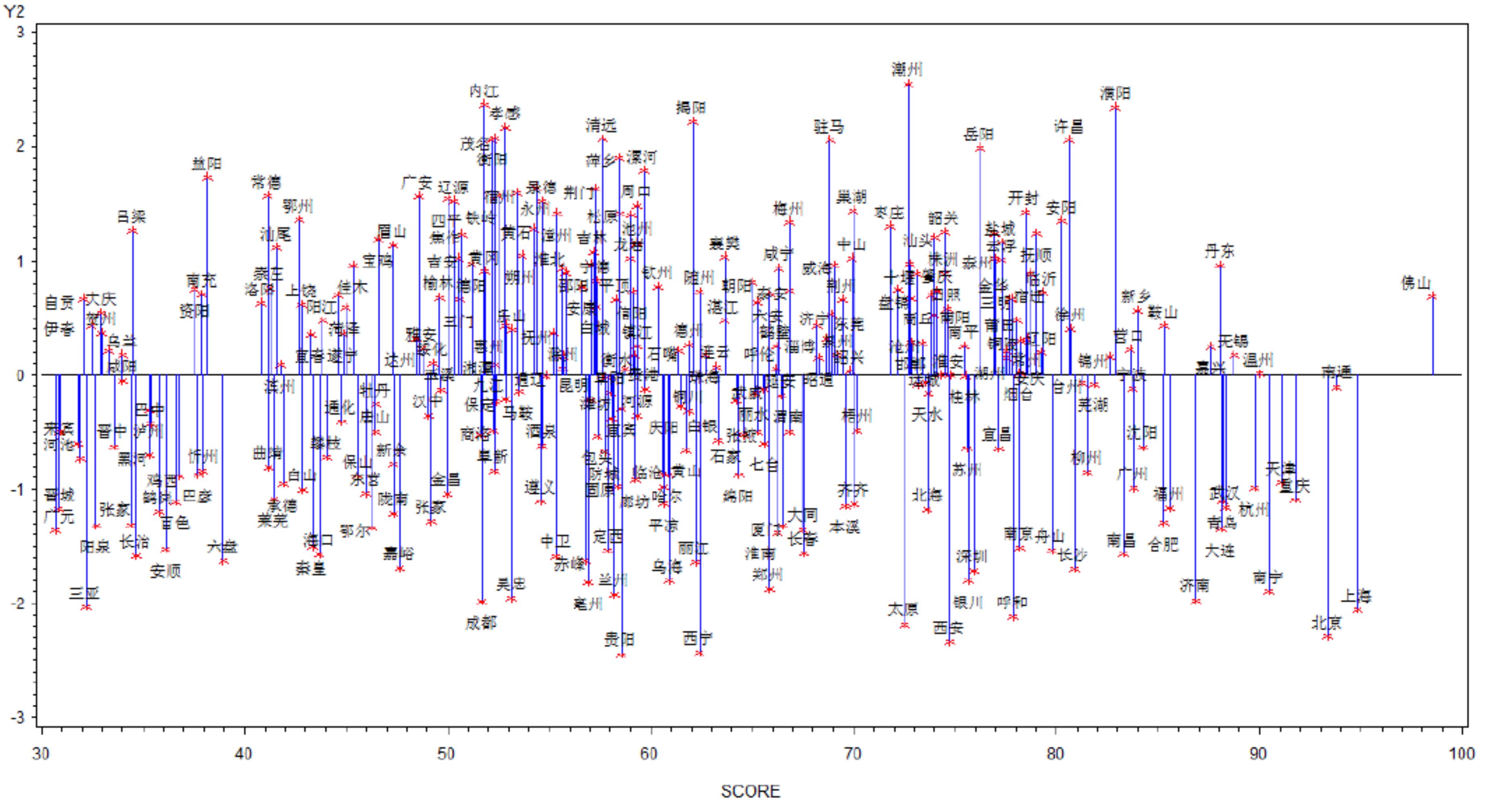}
\caption{\label{fig:city_rank} (City label in Chinese)Score plot for 284 participating China cities. The horizontal axis is a naive mapping of Z1(the first component value) into score($score=-11 \times Z1+61$), where Z1 is calculated via Eq.(\ref{eq:five}). The linear transformation of Z1 is to have our score within 30-100, which is comparable with official CEI that falls into the same range.}
\end{figure*}

A scatter plot of scores is also presented based upon the first principal component in Fig.\ref{fig:city_rank}. To make our scores directly comparable to official overall CEI, we apply a simple transformation to scale up raw Z1 values into expected CEI range. We do not label all 284 cities in English because of the intrinsic nature of the data collected by our Chinese colleagues. The score is formated so that higher value indicates better credit environment. For example, the city of Fo Shan, Shanghai, Nan Tong and Beijing are the top four in our study.

\begin{figure*}[htbp]
\includegraphics[width=170mm]{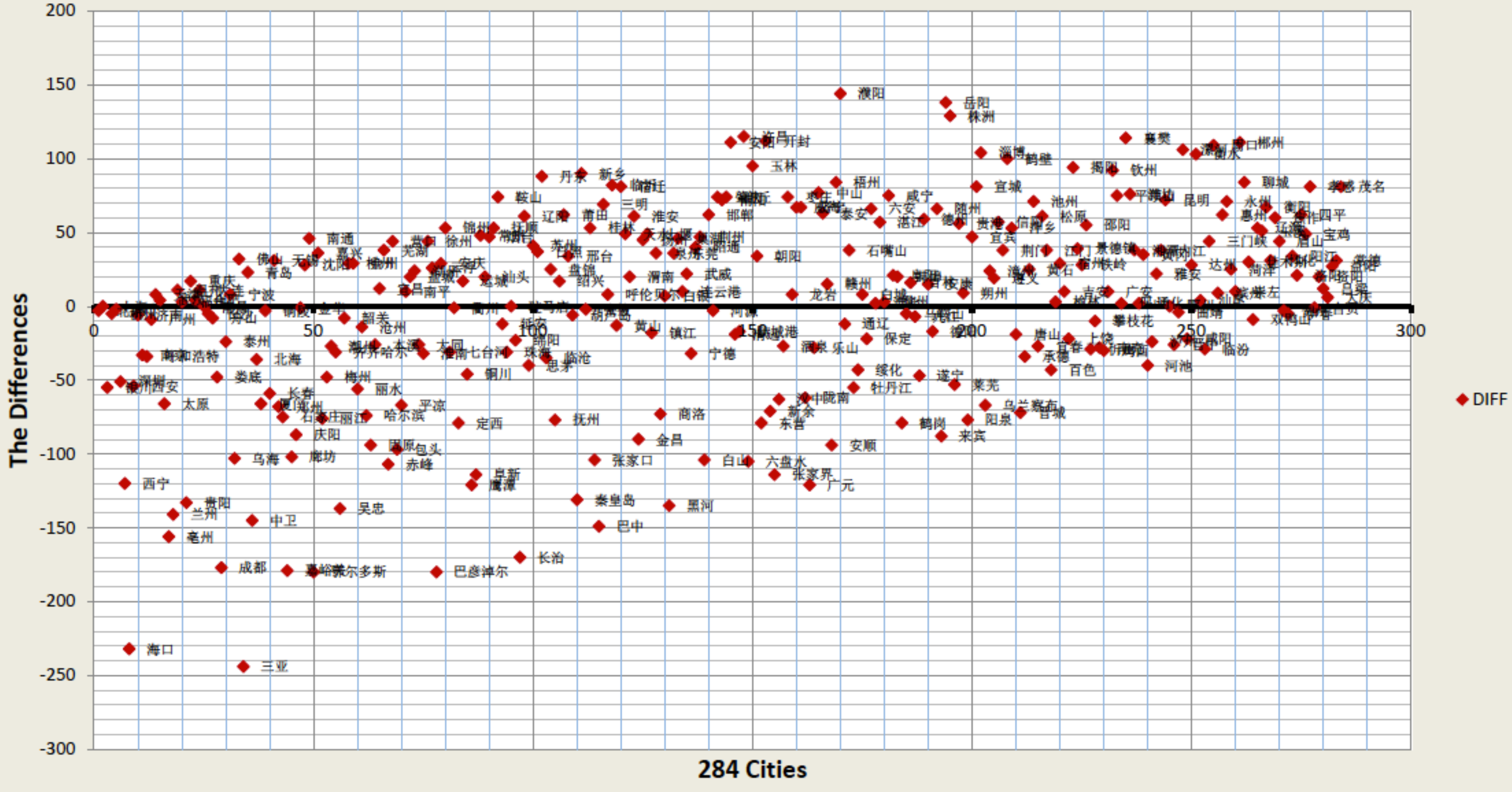}
\caption{\label{fig:city_scatterplot} (City label in Chinese)Rank difference plot for 284 participating China cities. The horizontal axis is official CEI values, and vertical axis is rank difference calculated via Eq.(\ref{eq:seven})}
\end{figure*}

To examine the validity of our analysis, a comparison with official overall CEI is needed. We visualize such comparison by plotting rank difference between our study and official overall CEI for all 284 participating mainland cities from China. The plot can be found in Fig.\ref{fig:city_scatterplot}. Again, the overall pattern of the data points are much more important than the city names those points correspond to, and we do not label the city names in English. In Figure.\ref{fig:city_scatterplot}, the horizontal axis represents official CEI values, and vertical axis demonstrates rank difference. The data points are clustered within $\pm$ 50 in rank difference for most of the 284 cities. 

\begin{eqnarray}
\mbox{rank difference} = &(\mbox{rank by official overall CEI}) \nonumber \\
                                        & - (\mbox{our rank by Z1})
\label{eq:seven}.
\end{eqnarray}

\begin{figure}[htbp]
\includegraphics[scale=0.3]{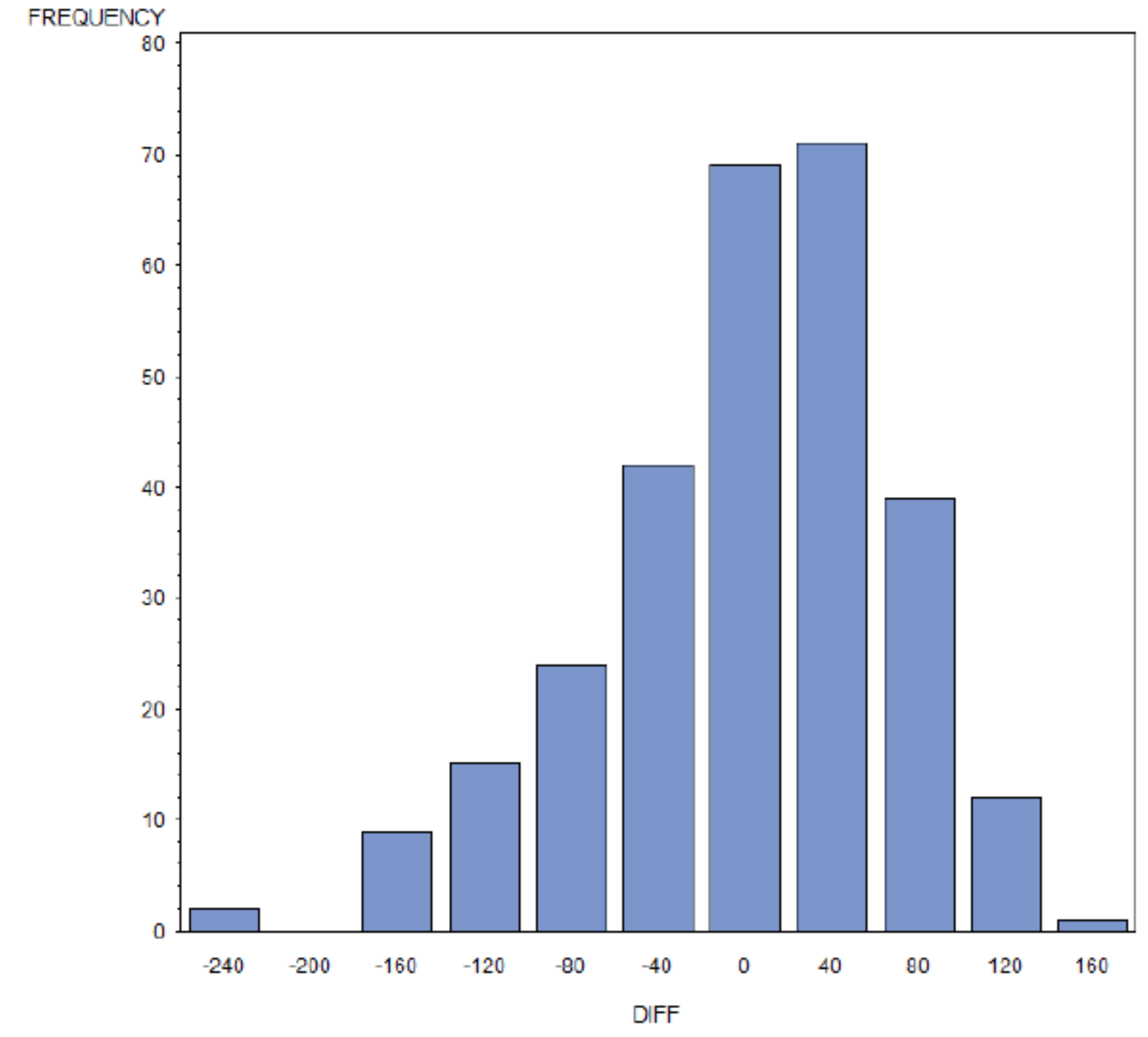}
\caption{\label{fig:city_diff_freq} Frequeny plot for rank difference.}
\end{figure}

The corresponding bar chart for rank difference is shown in Fig.\ref{fig:city_diff_freq}. To further examine the consistency between our studies and official CEI results, a paired difference test is performed. 

In statistics, a paired difference test is a type of location test that is used when comparing two sets of measurements to assess whether their population means differ. The Wilcoxon signed rank test is a non-parametric statistical hypothesis test, which is used when comparing two related samples, matched samples, or repeated measurements on a single sample to assess whether their population mean ranks differ. In this study, without the assumptions of normality and independence, the Wilcoxon signed rank test is used to test the significance of the differences between each two sets of measurements(i.e. calculated CEI and official CEI).

The results from Wilcoxon signed rank test can be found in Table.\ref{tab:table5}. We tend to accept the null hypothesis. That is to accept the two sets of measurements exhibit no difference. 

\begin{table}[htbp]
\centering
\begin{tabular}{|l|l|l|}
\hline
\multicolumn{3}{|c|}{Tests for Location: Mu0=0} \\
\hline
\mbox{ }\mbox{ }\mbox{ }\mbox{ }Test & -Statistics- & - - - - -p Value- - - - - \\ \hline
Student's t & \mbox{ }\mbox{ }\mbox{ }t \mbox{ }\mbox{ } \mbox{ } 0& Pr $>$ $\vert$t$\vert$  \mbox{ } \mbox{ } \mbox{ } \mbox{ }   1.0000 \\
\mbox{ } \mbox{ } Sign & \mbox{ }\mbox{ }\mbox{ }M  \mbox{ }\mbox{ }17& Pr $\geq$ $\vert$M$\vert$  \mbox{ } \mbox{ } \mbox{ }   0.0484 \\
Signed Rank & \mbox{ }\mbox{ }\mbox{ }S \mbox{ }\mbox{ }1699 & Pr $\geq$ $\vert$S$\vert$  \mbox{ } \mbox{ } \mbox{ } \mbox{ }  0.2108 \\ \hline
\end{tabular}
\caption{\label{tab:table5}The results from Wilcoxson signed rank test.}
\end{table}

\section{\label{sec:level1} Further Discussions}

The Wilcoxson signed rank test indicates a consistency achieved between our studies and official CEI results. Although we only use the first pincipal component to model our score, the result is already quite telling and explanatory. One possible potential study may be to include more pincipal components to see if any improvement could be obtained. We defer this topic to a future study.

We already know from Table.\ref{tab:table3} that the cumulative proportion if including first four principal components is up to 82\%, which is, under any circumstance, good enough to draw any strong conlusion. 

When official China cities CEI is calculated by China CEI research group, a weighted linear summation method is applied to include contributions from various sources. As stated earlier, determining weight is a quite subjective process, although advanced Analytic Hierarchy Process(AHP) method is employed to minimize the effects. The weight for credit launch index is $0.304075$, for enterprise risk management it is $0.141675$, $0.068975$ is determined for construction of credit reporting system, $0.1253$ for governmental credit minitoring, $0.1853$ for discredit/dishonesty, and $0.082475$ for credit education status. Although a weight of $0.0922$ is assigned for corporate experience, there is not any categorical CEI data(or categorical city rankings) by this index in CEI-2012, which is the reason why we only include six out of seven first-class indices in our studies.
 
The achieved consistency also implies those weights assigned by China CEI research group is reasonable. 

The readers may have noticed from Fig.\ref{fig:city_scatterplot} that there are apparently more cities sitting above ``rank difference = 0" reference line, which might indicate that official CEI exhibits, if any, an interesting trend to slightly ``overestimate" the goodness of China cities commercial credit environment. This interesting feature can also be easily captured from Fig.\ref{fig:city_diff_freq}, where center of the guassian-shape distribtuion is located in the right side of zero. A simple gussian-like fit shows the distribution is off to right by $\sim$ 20 city ranks. If it is there, this interesting trend would definitely render us a new direction to further adjust our assigned factor weights and/or improve our model in future. 

In addition, city rankings vary in the order of $\sim$ 30 in two different scenarios, which can be estimated by the standard deviation of our guassian fit. This model-dependent feature of rankings indicates that within small enough ranking range(let's say, 30), city rank is less indicative, and thus we cannot draw any clear conclusion as to whether one city is better/worse than another.

\section{\label{sec:level1} Summary}

CEI is designed to reflect various aspects of the comprehensive urban commercial credit environment. For pratical reasons, CEI needs to take into account the feasibility of credit information collecting and the authority of data. CEI is constructed based on multiple factors(called index), which are categorized into three major classes. We present an independent statistical study using categorical city rankings by each of the six first-class indices via principal component analysis techniques. We are able to construct, in parallel, an independent CEI that reflects same features of commercial credit environment among all 284 participating mainland cities in China. The consistency between our studies presented in prior section of this paper and official CEI results indicates CEI is indeed a convincing and powerful tool to quantify and evaluate the status quo of commercial environment among mainland cities in China. 

Based on our studies, the factor of government-level credit monitoring services and credit education status are the two most influential indices in modeling CEI. The extraction of important factors serves as a guideline for future policy making so as to build a better overall commercial credit environment across major mainland cities in China.

\vspace{6pt}

\section{\label{sec:level1} Acknowledgement}

We would like to thank China CEI research group for giving us the opportunity to participate in China cities CEI studies and providing data for our analysis. We appreciate valuable comments from CEI research group and intriguing discussions with professor Junyue Lin.

\nocite{*}

\bibliography{apssamp}

\end{document}